\documentclass[aaspp4,11pt,preprint,apjfonts]{emulateapj}

\newcommand{\um}{$\mu$m}

\shorttitle{Mid-infrared observations of NGC4038/4039}
\shortauthors{Snijders et al.}

\begin{document}

\title{Subarcsecond resolution mid-infrared observations of super star clusters in the Antennae (NGC4038/4039)\altaffilmark{1}}

\author{L. Snijders\altaffilmark{2},
P.~P. van der Werf\altaffilmark{2},
B.~R. Brandl\altaffilmark{2}, 
S. Mengel\altaffilmark{3}, 
D. Schaerer\altaffilmark{4},
Z. Wang\altaffilmark{5}}

\altaffiltext{1}{Based on observations collected at the
European Southern Observatory, Paranal, Chile, under programme no. 075.B-0791(A)} 
\altaffiltext{2}{Leiden Observatory, PO Box 9513, 2300 RA Leiden, The Netherlands}
\altaffiltext{3}{European Southern Observatory, Karl-Schwarzschild-Str. 2, 85748 Garching, Germany}
\altaffiltext{4}{Observatoire de Gen\`eve, 51, Ch. des Maillettes, CH-1290 Sauverny, Switzerland and\\
Laboratoire d'Astrophysique (UMR 5572), Observatoire Midi-Pyr\'en\'ees, 14 Avenue E. Belin, F-31400 Toulouse, France}
\altaffiltext{5}{Smithsonian Astrophysical Observatory, Cambridge, MA 02138, USA}


\begin{abstract}

In this letter we present ground-based subarcsecond mid-infrared
imaging and spectroscopy of young super star clusters in the overlap
region of the merging galaxies NGC4038/4039 (the Antennae) obtained
with the VLT Imager and Spectrometer for mid-Infrared (VISIR). With
its unprecedented spatial resolution VISIR begins to resolve the
H\,{\sc ii}/PDR complexes around the star-forming regions for the
first time. In the N-band spectra of two young star clusters
unexpectedly low polycyclic aromatic hydrocarbon (PAH) emission is
observed, compared to what is seen with the Infrared Space Observatory
(ISO) and with the Spitzer Space Telescope. We conclude that a large
fraction of the PAH emission cannot directly be associated with the
super star clusters, but originate from an extended region of at least
215 pc radius around the clusters.  In the distribution of [Ne\,{\sc
ii}] 12.81 \um ~emission a highly obscured cluster is discovered that
does not have an optical or near-infrared counterpart.

\end{abstract}

\keywords{galaxies: individual (NGC4038/4039) --- galaxies: star clusters --- ISM: dust/PAH}


\section{Introduction}

Starburst galaxies experience a phase of rapid evolution. The
rate at which their gas reservoir is turned into stars cannot be
maintained for long, which makes the starburst phase by definition a
transient one. The resulting stellar clusters make
starbursts unique laboratories for the study of star formation,
stellar populations and the evolution of galaxies as a whole.

Since the earliest stages of massive star formation are generally
heavily enshrouded by dust, infrared observations are essential to
detect the youngest stellar populations and provide various diagnostic
features to study the properties of the interstellar matter (ISM) and
the underlying stellar population. Recently, a new generation of
ground- and space-based instruments working at mid-infrared
wavelengths has become operational, giving this field of research a
large impulse.

The Antennae (NGC4038/4039, Arp244) is the nearest major merger of two
large spirals \citep{Toomre:1972}. Since the beginning of the
interaction the system went through several episodes of violent star
formation of which the last one is probably still ongoing
\citep{Vigroux:1996}. The resulting (super) star clusters have been
studied extensively throughout the electromagnetic spectrum
\citep{Whitmore:2005, Wang:2004, Gilbert:2000}. Radio and mid-infrared
observations showed that the region between the two spirals (the
overlap region) hosts spectacular obscured star formation. The
brightest mid-infrared component produces 15\% of the total 15 \um
~luminosity of the entire system \citep{Mirabel:1998,
Hummel:1986}. This region is covered by a prominent dustlane, and can
be associated with a faint, red source in the HST images \citep[number
80 of][hereafter WS95]{Whitmore:1995}, illustrating how optical data
alone are insufficient to identify and study the youngest star-forming
regions. An ISOCAM spectrum of this region shows strong fine-structure
emission lines ([Ne\,{\sc ii}] and [Ne\,{\sc iii}]) and pronounced
emission from polycyclic aromatic hydrocarbons (PAHs)
\citep{Mirabel:1998}.

We use the recently commissioned VLT Imager and Spectrometer for
mid-Infrared (VISIR) \citep{Lagage:2004} at the Very Large Telescope
(VLT) of the European Southern Observatory (ESO) to study the most
luminous super star clusters and the surrounding matter in the
Antennae overlap region in detail.


\section{Observations and data reduction}
The southern part of the overlap region of NGC4038/39 was observed
with VISIR on 2005 April 17 -- 20. VISIR offers a large set of imaging
filters covering the 8 -- 13 and the 17 -- 24 micron range of which we
used four (see Table~\ref{sizes}), all at the largest pixel scale of
0\farcs127/pixel, resulting in a 32.5\arcsec $\times$ 32.5\arcsec
~field-of-view. To correct for the strong and highly variable
atmospheric background chopping and nodding was applied, with a
chopper throw of 14\arcsec. The total on-source integration time was
60 minutes in both the PAH and PAH reference filter, 120 minutes in
the [Ne\,{\sc ii}] filter and 15 minutes in the [Ne\,{\sc ii}]
reference filter. The latter observations had to be aborted
prematurely because of strong winds. Between individual nod cycles the
telescope was given a random jitter offset to correct for array
artefacts. Standard stars \citep[HD99167 and HD157999,][]{Cohen:1999}
were observed before and after each target observation for flux
calibration.

Additionally, low-resolution N-band spectra were obtained for the two
brightest sources in the overlap region simultaneously (R $\approx$
185 -- 390, 0\farcs75 slit). Four spectral settings overlapping by at
least 15\% were used to cover the full N-band ($\lambda_{\rm centr}$ =
8.8, 9.8, 11.4 and 12.4 \um). The total on-source integration time was
30 minutes for each spectroscopic setting. Since the images showed no
large-scale diffuse emission, chopping and nodding on-slit was applied
with a 10\arcsec ~chopper throw. Early-type stars were observed before
and after each target observation for flux calibration (HD115892 and
HD135742).

IRAF\footnote[1]{Image Reduction and Analysis Facility (NOAO, National
Optical Astronomy Observatories)} routines and customized IDL scripts
were used for data reduction. Subtracting the chopping and nodding
pairs removes most of the sky and telescope background. The resulting
images were shifted and coadded. Wavelength calibration and removal of
the curvature in the spectroscopic data was done by tracing
skylines. Spectra were extracted in a 1\farcs27 ~aperture. Absolute
flux calibration was obtained by normalizing the standard star spectra
to VISIR narrow-band fluxes. The accuracy of the flux calibration is
20\% both for imaging and spectroscopy.

\begin{figure}
\epsscale{1.18}
\plotone{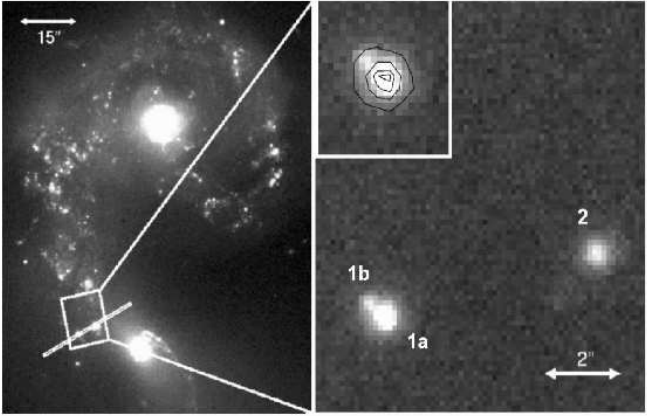}
\caption{Left panel: SoFI Ks-band (2.16 \um). The VISIR slit is shown in 
white, north is up, east is left. Right panel: VISIR [Ne\,{\sc ii}] 12.81 \um ~of 
the overlap region, inserted: source 1a and 1b in the [Ne\,{\sc ii}] 12.81 \um 
~filter with the contours of PAH 11.25 \um ~filter emission overlayed. 
\label{Ant}}
\end{figure}

\begin{figure}
\epsscale{1.18}
\plotone{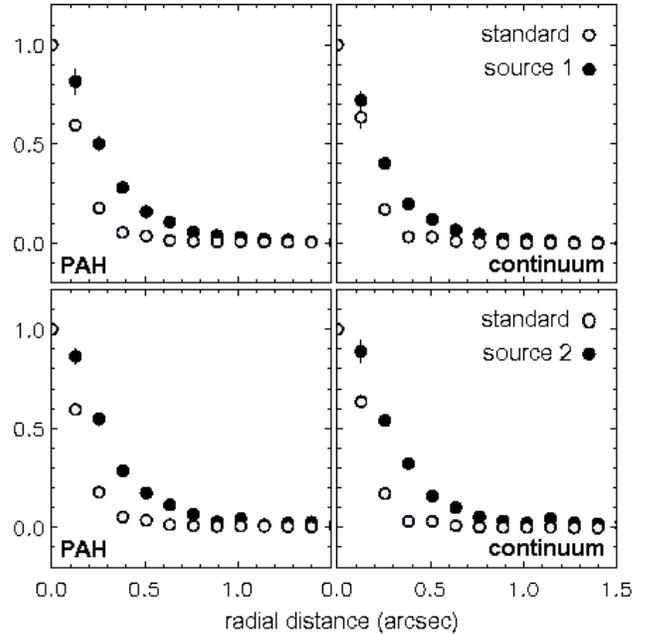}
\caption{Normalized radial profiles of source 1 (upper panels) and 2
(lower panels) in the PAH (left panels) and reference continuum
filters (right panels). Filled circles show the profile of source 1/2,
open circles that of all standard stars combined. Error bars are
plotted, but are smaller than the plotting symbol, expect for the
points at smallest radial distance.\label{radprof}}
\end{figure}

\begin{figure}
\epsscale{1.18}
\plotone{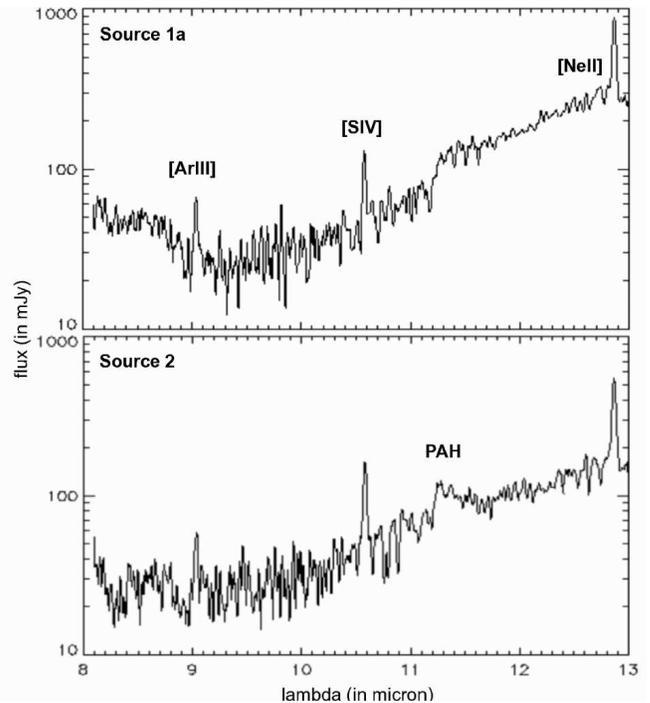}
\caption{Upper panel: low resolution N-band spectrum of source 1a. The apparently 
enhanced noise from 9 -- 10 \um ~is a result of the log scale of this plot. The observed fine-structure line ratio [S\,{\sc iv}]/[Ne\,{\sc ii}] is
higher than obtained within the larger apertures of ISO and Spitzer,
indicating that VISIR probes regions of higher excitation closing in
on the location of most intense star formation.
Lower panel: idem source 2. \label{Ant2}}
\end{figure}


\section{Results}

\begin{deluxetable}{llllcl}
  \tabletypesize{\footnotesize} \tablecaption{Cluster photometry and sizes \label{sizes}}
  \tablewidth{0pt} 
\tablehead{ \colhead{filter} &
            \colhead{$\lambda_{\rm centr}$} & 
	    \colhead{$\Delta\lambda$} &
	    \colhead{seeing\tablenotemark{a}} & 
	    \colhead{flux density\tablenotemark{b}} & 
	    \colhead{size\tablenotemark{c}}\\
	     &
            \colhead{(\um)} & 
            \colhead{(\um)} & 
	     & 
	    \colhead{(mJy)} & 
	    \colhead{(pc)}
 }

 \startdata \multicolumn{6}{c}{Source 1} \\ 
  PAH                            & 11.25  & 0.59 & 0\farcs33 $\pm$ 0\farcs03 & 98 $\pm$ 20                          & 88 $\pm$ 14  \\
  PAHref                         & 11.88  & 0.37 & 0\farcs32 $\pm$ 0\farcs01 & 133 $\pm$ 27                         & 80 $\pm$ 14  \\ 
  $[$Ne\,{\sc ii}$]$ref\tablenotemark{d} & 12.27  & 0.21 & 0\farcs46 $\pm$ 0\farcs01 & 203 $\pm$ 41                 & -- \\
  $[$Ne\,{\sc ii}$]$                     & 12.81  & 0.18 & 0\farcs34 $\pm$ 0\farcs02 & 373 $\pm$ 75\tablenotemark{e}& 93 $\pm$ 20  \\ 
  & & & & \\
  \multicolumn{6}{c}{Source 2} \\ 
  PAH                            & 11.25  & 0.59 & 0\farcs33 $\pm$ 0\farcs03 & 74 $\pm$ 15          & 76 $\pm$ 9    \\
  PAHref                         & 11.88  & 0.37 & 0\farcs32 $\pm$ 0\farcs01 & 84 $\pm$ 17          & 71 $\pm$ 9    \\ 
  $[$Ne\,{\sc ii}$]$ref\tablenotemark{d} & 12.27  & 0.21 & 0\farcs46 $\pm$ 0\farcs01 & 114 $\pm$ 23 & -- \\
  $[$Ne\,{\sc ii}$]$                     & 12.81  & 0.18 & 0\farcs34 $\pm$ 0\farcs02 & 210  $\pm$42 & 84 $\pm$ 10    \\ 
  \enddata

\tablenotetext{a}{mean of FWHM of PSF of all standard stars observed before and after target observations}
\tablenotetext{b}{total flux density, sum of compact and broad component}
\tablenotetext{c}{FWHM of the broad component of the best fit}
\tablenotetext{d}{low SNR; observations aborted prematurely because of strong wind}
\tablenotetext{e}{sum of sources 1a and 1b, one-fifth can be associated with 1b}

\end{deluxetable}

The narrow-band PAH 11.25 \um ~and reference images show two bright
resolved sources separated by 5\farcs8 ~($\approx$ 600 pc in
projection at a distance of 21 Mpc\footnote[2]{Note that the distance
of the Antennae is under debate. Recently a lower distance of 13.8
$\pm$ 1.7 Mpc was found \citep{Saviane:2004}, which would affect the
values derived here.}, assuming a Hubble constant of 70
km~s$^{-1}$~Mpc$^{-1}$; at this distance 1\arcsec ~corresponds to 105
pc) in the southern part of the overlap region. These sources
correspond to the two strong peaks identified in the overlap region in
ISO images \citep{Mirabel:1998}. The eastern source is the counterpart
of the faint, very red cluster 80 in HST images (WS95) and a bright
cluster in the near-infrared \citep[ID$_{\rm WIRC}$
157,][]{Brandl:2005}. The western source can be associated with a very
crowded, much less obscured, bright cluster complex consisting of the
star clusters WS95-86 to WS95-90 plus a number of close neighbours and
with another bright near-infrared cluster, ID$_{\rm WIRC}$ 136. It is
remarkable how similar the eastern and western source are in shape and
brightness in the mid-infrared, while these regions appear very
different at shorther wavelengths. Given the similar ages for these
clusters \citep[both $<$ 6 Myr][]{Gilbert:2000, Mengel:2005}, the
differences in appearance at the near-infrared and the optical
indicate strongly varying physical conditions and/or extinction in
these star-forming regions.

In the [Ne\,{\sc ii}] filter image the eastern source breaks up into two
distinct sources, which are separated by 0\farcs54 ($\approx$ 60 pc in
projection, Fig.~\ref{Ant}: source 1a \& 1b). Relative astrometry
strongly suggests that source 1a can be associated with the bright
near-infrared cluster ID$_{\rm WIRC}$ 157 and with the optical source
WS95-80 mentioned before. This was determined by matching source 2
through all wavelength bands: the peak in the mid-infrared data was
accurately matched with the well-defined peak in the
near-infrared. The morphology of the cluster complex in the
near-infrared was in turn precisely overlayed with the optical
data. The fainter second source 1b, emitting one-fifth of the total
emission in the [Ne\,{\sc ii}] filter (continuum plus [Ne\,{\sc ii}]
line flux), has not been observed before. Neither in the optical nor
in the near-infrared possible counterparts are identified. At shorter
wavelengths this cluster is hidden by extinction and at longer
wavelengths most instruments lack spatial resolution to separate the
two clusters.

Under the assumption that source 1b is as 1a younger than 6 Myr, that
source 1b has a similar spectral energy distribution and using the
values for the Ks-band magnitude and extinction as published \citep[Ks
= 14.59 and $A_{\rm V}$ is 4.23, Mengel et al. 2005; assuming that
$A_{\rm V}/A_{\rm K}$ $\approx$ 9.3,][]{Mathis:1990}, the minimal
extinction necessary to account for the observations is $A_{\rm K}$
$>$ 7.8 ($A_{\rm V}$ $>$ 72) of source 1b. Although the exact
intrinsic Ks-band flux of source 1b is not straightforward to
determine, it is obvious that it is extinguished by an enormous amount
of obscuring matter. The presence of this cluster compromises efforts
to model the emission from this region with one coeval stellar
population. It might for example partly explain the discrepancy
between the mid-infrared luminosity measured in this region by
ISO \citep[15\% of the total 15\um ~flux of the entire system,
][]{Mirabel:1998} and the far lower expected mid-infrared luminosity
as reconstructed from the properties of the underlying stellar
population derived from near-infrared and optical data \citep[3.6\% of
the total 15\um ~flux,][]{Mengel:2005}.

Fig.~\ref{radprof} shows that both source 1 and 2 are spatially
resolved. The radial profiles of the sources are well fitted by a
combination of two Gaussians, one compact component and a more
extended component, both contributing roughly 50\% to the total
flux. The full-width at half maximum (FWHM) of the broad component for
the best fit varies from 0\farcs85 -- 1\farcs0 for source 1 and
slightly smaller, 0\farcs75 -- 0\farcs90, for source 2. This
corresponds to a size of 90 -- 105 pc and 80 -- 95 pc respectively,
similar to sizes observed for giant molecular
clouds. Table~\ref{sizes} lists the size of the broad component and
flux density ${\rm F_\nu}$ for both sources in all filters.

The spatial distribution along the slit in the spectroscopic data
shows two bright, slightly resolved sources without any obvious more
extended emission. The extracted N-band spectra show the typical
signatures of a region of massive star formation: fine structure lines
($\rm [Ne II]$, [S\,{\sc iv}] and faint [Ar\,{\sc iii}]) and a rising
continuum characteristic for thermal emission by dust.  Near-infrared
data suggest that these clusters are very young, which is supported by
the presence of strong [S\,{\sc iv}] emission. Line fluxes and
equivalent widths (EWs) are given in Table \ref{fluxes}.

Besides obvious similarities in the general shape of the N-band
spectra (Fig.~\ref{Ant2}) there are some remarkable differences from
previous observations of the same regions by ISO \citep[Fig.1
in][]{Mirabel:1998} and Spitzer (Brandl et al, in preparation). The
most striking contrast is the strength of the PAH features. Where the
ISO and Spitzer spectra of the region around complex 1 show a
pronounced 11.3 $\mu$m PAH-feature, the VISIR spectrum only shows a
slight indication of it. 


\section{PAH emission}

\begin{deluxetable}{llcl}
  \tabletypesize{\footnotesize} \tablecaption{Line fluxes from spectroscopy \label{fluxes}}
  \tablewidth{0pt} 
\tablehead{ \colhead{ Species} &
            \colhead{$\lambda_{\rm rest}$} & 
	    \colhead{line flux} & 
            \colhead{EW} \\
	    &
            \colhead{(\um)} & 
	    \colhead{(10$^{-14}$ erg~s$^{-1}$~cm$^{-2}$)} & 
	    \colhead{(\um)} 
 }

 \startdata

\multicolumn{4}{c}{Source 1a} \\    
            $[$Ar\,{\sc iii}$]$ &  8.99138         & 3.9  $\pm$ 0.9    & 0.033 $\pm$ 0.009   \\
            $[$S\,{\sc iv}$]$   & 10.5105          & 7.6  $\pm$ 0.9    & 0.057 $\pm$ 0.011   \\
	    PAH                 & 11.3             & 14.0 $\pm$ 3.8    & 0.057 $\pm$ 0.017   \\
            $[$Ne\,{\sc ii}$]$  & 12.8136          & 49.6 $\pm$ 6.2    & 0.092 $\pm$ 0.018   \\
			 &		    & 	  &	\\
\multicolumn{4}{c}{Source 2} \\   
            $[$Ar\,{\sc iii}$]$ &  8.99138          & 4.2  $\pm$ 0.7   & 0.051 $\pm$ 0.011 \\
            $[$S\,{\sc iv}$]$    & 10.5105          & 10.3 $\pm$ 2.3   & 0.078 $\pm$ 0.019 \\
	    PAH                  & 11.3             & 25.0 $\pm$ 4.2   & 0.127 $\pm$ 0.029 \\
            $[$Ne\,{\sc ii}$]$   & 12.8136          & 30.2 $\pm$ 4.5   & 0.111 $\pm$ 0.023 \\

  \enddata

\end{deluxetable}

One of us (BRB) is leading a GTO program to observe the Antennae with
Spitzer-IRS. The results will be published in a subsequent paper
(Brandl et al. in preparation). These observations include
observations of both complex 1 and source 2. 

Comparison of the VISIR spectrum of source 1a published here with the
Spitzer spectrum of the same region shows considerable differences. The
continuum flux density measured by VISIR is roughly three-quarters of
that observed by Spitzer. Of the total emission in the 11.3 $\mu$m PAH
feature VISIR measures a much smaller fraction, only approximately
one-quarter of the emission seen in the Spitzer spectrum. This large
difference indicates that the region of PAH emission is significantly
more extended than that of continuum emission.

Because of its wider slit Spitzer is expected to collect more flux
than VISIR in case of extended sources (4\farcs7 and 0\farcs75
respectively). Indeed, VISIR finds a lower flux density than Spitzer at
all wavelengths, but the relative fraction is different for the
various spectral features. The EW of the 11.3 $\mu$m PAH feature found
with VISIR is over three times smaller than the EW measured within the
larger apertures of both Spitzer and ISO \citep{Mirabel:1998}. The
total flux in the PAH feature in the VISIR spectrum of source 1a is
(1.4 $\pm$ 0.4) $\cdot$ $10^{-13}$~erg~s$^{-1}$~cm$^{-2}$. The remaining
three-quarters of PAH emission that is ``missing'' compared to the Spitzer
data originate from a more extended region. If we assume that this (4.2
$\pm$ 0.8) $\cdot$ $10^{-13}$~erg~s$^{-1}$~cm$^{-2}$ of PAH emission is
homogeneously distributed and has a surface brightness just below the
detection limit of our data (2$\sigma$ is 2.56 $\cdot$
$10^{-14}$~erg~s$^{-1}$~cm$^{-2}$~arcsec$^{-2}$), the emission must
originate from a region of at least 215 $\pm$ 25 pc radius.

\begin{figure}
\begin{center}
\epsscale{0.7}
\plotone{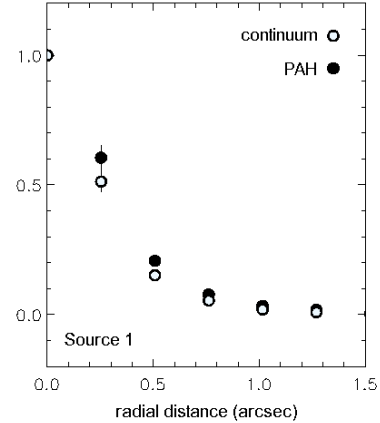}
\caption{Normalized radial profile of source 1 in the PAH (filled
circles) and reference continuum filters (open circles). The profile
in the PAH filter drops off slightly softer than the continuum
profile. Error bars are plotted, but are smaller than the plotting
symbol, expect for the points at smallest radial distance.
\label{radprof2}}
\end{center}
\end{figure}

While this faint emission cannot directly be detected in our data,
indications for more extended PAH emission are also found by close
examination of the radial profiles in both the PAH and the reference
continuum filter where a slightly softer drop-off of the profile in
the PAH filter is found (see Fig.~\ref{radprof2}).

These results indicate that an extended PAH emission component below the
VISIR surface brightness detection limit is present that accounts for
the discrepancy of the VISIR with the ISO and Spitzer data. Spitzer
IRAC images indeed show indications of widespread PAH emission
\citep{Wang:2004}.

For source 2 there are no ISO data available, but comparing the VISIR
and Spitzer data leads in a similar way to the conclusion that
extended PAH emission must be present. The VISIR spectrum of this
source shows a more pronounced PAH feature. The EW is roughly two
times smaller than that measured in the Spitzer spectrum and of the
total PAH flux one-third is measured by VISIR. Compared to source 1a
we measure a smaller fraction of the total continuum flux, only
two-thirds of the continuum flux density measured by Spitzer. The
smaller fraction of the continuum measured and the smaller discrepancy
between the VISIR and Spitzer EWs reflects the extended structure of
the underlying rich cluster complex. In contrast to the case with the
relatively simple morphology of source 1, here a smaller fraction of
the spatially extended structure is covered by the narrow VISIR
slit. Although the morphology of source 2 is simple in our imaging
data, the fact that we only measure half of the continuum compared to
Spitzer indicates that a more extended component must be present below
the VISIR detection limit. Taking this into account the PAH emission
must still originate from a more extended region than the continuum.

These results indicate that VISIR is resolving the H\,{\sc ii}/PDR
complexes. While the ionized gas and continuum emission, and thus the
hard radiation field of the most recent star formation, are much more
concentrated, a large fraction of the PAH emission originates from a
region more extended than 215 pc around the clusters. This could be
explained by the presence of a (slightly) more evolved, less massive
and more widespread population of field stars by which the PAH
molecules are excited. This population can consist of the original,
pre-merger disk stars or can be the result of dissolution of clusters
formed in previous episodes of star formation. The timescale of the
latter process is estimated to be a few 10 Myr in the Antennae
\citep{Mengel:2005}. In this scenario a large fraction of the PAH
emission is not directly associated with the currently observed super
star clusters and therefore does not trace the very most recent star
formation. PAH emission has been detected in less UV-rich environments
before \citep{Uchida:1998}. \cite{Li:2002} argue that it does not
necessarily take hard UV radiation to excite PAHs and produce the
observed emission bands in the 6--12 \um ~range. Another possible
explanation is the destruction of PAH molecules by hard UV photons up
to large distances from the ionizing clusters, which is also claimed
to be seen in other starburst galaxies \citep{Tacconi:2005,
Beirao:2006}. However, source 2 shows the strongest PAH emission and
has the highest [S\,{\sc iv}]/[Ne\,{\sc ii}] ratio (0.15 $\pm$ 0.02
and 0.34 $\pm$ 0.04 for sources 1a and 2 respectively). This
significantly higher ratio, indicative of a harder radiation field,
would argue against the PAH destruction scenario in this case.

\section{Future work}

The VISIR N-band observations of the Antennae overlap region presented
in this letter show the power of a ground-based mid-infrared
instrument mounted on a large telescope. Due to the unprecedented
spatial resolution in the mid-infrared, which space observatories
lack, VISIR is an excellent tool to use in detailed studies of nearby
starburst galaxies. Structures on the scale of giant molecular clouds
(80 -- 105 pc) are resolved, each hosting one or a number of star
clusters.

In order to further investigate the excitation of PAHs in starburst
environments, we are currently studying a number of starburst galaxies
with a range in age and luminosity.

\acknowledgments
We thank the Paranal Observatory Team for their support.



\begin{thebibliography}{}

  \bibitem[Beir\~ao et al., 2006]{Beirao:2006} Beir\~ao, P., Brandl, B.~R., Devost, D., Smith, J.~D., Hao, L. \& Houck, J.~R., 2006, ApJ 643, 1 

  \bibitem[Brandl et al., 2005]{Brandl:2005} Brandl, B.~R. et al. 2005, ApJ 635, 280

  \bibitem[Cohen et al., 1999]{Cohen:1999} Cohen, M., Walker, R.~G., Carter, B., Hammersley, P., Kidger, M. \& Noguchi, K., 1999, AJ 117, 1864

  \bibitem[Gilbert et al., 2000]{Gilbert:2000} Gilbert, A.~M. et al. 2000, ApJ 533, L57

  \bibitem[Hummel \& Van der Hulst, 1986]{Hummel:1986} Hummel, E. \& Van der Hulst, J.~M., 1986, A\&A 155, 151

  \bibitem[Lagage et al., 2004]{Lagage:2004} Lagage, P.~O. et al. 2004, ESO Messenger No. 117, 12

  \bibitem[Li \& Draine, 2002]{Li:2002} Li, A. \& Draine, B.~T., 2002, ApJ 572, 232

  \bibitem[Mathis, 1990]{Mathis:1990} Mathis, J.~S., 1990, ARA\&A 28, 37

  \bibitem[Mengel et al., 2005]{Mengel:2005} Mengel, S., Lehnert, M.~D., Thatte, N. \& Genzel, R., 2005, A\&A 443, 41

  \bibitem[Mirabel et al., 1998]{Mirabel:1998} Mirabel, I.~F. et al. 1998, A\&A 333, L1

  \bibitem[Neff \& Ulvestad, 2000]{Neff:2000} Neff, S.~G. \& Ulvestad, J.~S., 2000, AJ 120, 670

  \bibitem[Rubin et al., 1970]{Rubin:1970} Rubin, V.~C., Ford, W.~K.~J. \& D'Odorico, S., 1970, ApJ 160, 801

  \bibitem[Saviane et al., 2004]{Saviane:2004}	Saviane, I., Hibbard, J.~E. \& Rich, M.~R., 2004, AJ 127, 660

  \bibitem[Tacconi-Garman et al., 2005]{Tacconi:2005} Tacconi-Garman, L.~E., Sturm, E., Lehnert, M., Lutz, D., Davies, R.~I \& Moorwood, A.~F.~M., 2005, A\&A 432, 91

  \bibitem[Toomre \& Toomre, 1972]{Toomre:1972} Toomre, A. \& Toomre, J., 1972, ApJ 178, 623

  \bibitem[Uchida et al., 1998]{Uchida:1998} Uchida, K.~I., Sellgren, K. \& Werner, M., 1998, ApJ 493, 109

  \bibitem[Vigroux et al., 1996]{Vigroux:1996} Vigroux, L. et al. 1996, A\&A 315, L93

  \bibitem[Wang et al., 2004]{Wang:2004} Wang, Z. et al. 2004, ApJ 154, 193

  \bibitem[Whitmore \& Schweizer, 1995]{Whitmore:1995} Whitmore, B.~C. \& Schweizer, F., 1995, AJ 109, 960

  \bibitem[Whitmore et al., 1999]{Whitmore:1999} Whitmore, B.~C., Zhang, Q., Leitherer, C., Fall, M.~S., Schweizer, F. \& Miller, B.~W., 1999, AJ 118, 1551

  \bibitem[Whitmore et al., 2005]{Whitmore:2005} Whitmore, B.~C. et al. 2005, AJ 130, 2104

\end{thebibliography}
\end{document}